\documentclass[aps,prb,reprint,superscriptaddress]{revtex4-1}

\usepackage{graphicx}
\usepackage{amsmath}
\usepackage{textcomp} 
\usepackage{xcolor}
\usepackage{ulem}

\newcommand{\YIG}{Y$_\mathrm{3}$Fe$_\mathrm{5}$O$_\mathrm{12}$}
\newcommand{\g}{$g_\mathrm{eff}^{\uparrow \downarrow}$}

\begin{document}

\title{Thickness and power dependence of the spin-pumping effect in \YIG/Pt heterostructures measured by the inverse spin Hall effect}

\author{M.~B.~Jungfleisch}
\email{jungfleisch@physik.uni-kl.de}
\affiliation{Fachbereich Physik and Landesforschungszentrum OPTIMAS, Technische Universit\"at Kaiserslautern, 67663
Kaiserslautern, Germany}

\author{A.~V.~Chumak}
\affiliation{Fachbereich Physik and Landesforschungszentrum OPTIMAS, Technische Universit\"at
Kaiserslautern, 67663 Kaiserslautern, Germany}

\author{A.~Kehlberger}
\affiliation{Institute of Physics, Johannes Gutenberg-University Mainz, 55099 Mainz, Germany}

\author{V.~Lauer}
\affiliation{Fachbereich Physik and Landesforschungszentrum OPTIMAS, Technische Universit\"at
Kaiserslautern, 67663 Kaiserslautern, Germany}

\author{D.~H.~Kim}
\affiliation{Department of Materials Science and Engineering, MIT, Cambridge, MA 02139, USA}
 
\author{M.~C.~Onbasli}
\affiliation{Department of Materials Science and Engineering, MIT, Cambridge, MA 02139, USA}

\author{C.~A.~Ross}
\affiliation{Department of Materials Science and Engineering, MIT, Cambridge, MA 02139, USA}

\author{M.~Kl\"aui}
\affiliation{Institute of Physics, Johannes Gutenberg-University Mainz, 55099 Mainz, Germany}

\author{B.~Hillebrands}
\affiliation{Fachbereich Physik and Landesforschungszentrum OPTIMAS, Technische Universit\"at
Kaiserslautern, 67663 Kaiserslautern, Germany}

\date{\today}

\begin{abstract}

The dependence of the spin-pumping effect on the yttrium iron garnet (\YIG, YIG) thickness detected by the inverse spin Hall effect (ISHE) has been investigated quantitatively.
Due to the spin-pumping effect driven by the magnetization precession in the ferrimagnetic insulator \YIG\ film a spin-polarized electron current is injected into the Pt layer. This spin current is transformed into electrical charge current by means of the ISHE. An increase of the ISHE-voltage with increasing film thickness is observed and compared to the theoretically expected behavior.  
The effective damping parameter of the YIG/Pt samples is found to be enhanced with decreasing \YIG\ film thickness.
The investigated samples exhibit a spin mixing conductance of \g~=~(7.43$ \pm$0.36)$ \times 10^{18}$ m$^{-2}$ and a spin Hall angle of $\theta_\mathrm{ISHE} = 0.009 \pm 0.0008$.
Furthermore, the influence of nonlinear effects on the generated voltage and on the Gilbert damping parameter at high excitation powers are revealed. It is shown that for small YIG film thicknesses a broadening of the linewidth due to nonlinear effects at high excitation powers is suppressed because of a lack of nonlinear multi-magnon scattering channels. We have found that the variation of the spin-pumping efficiency for thick YIG samples exhibiting pronounced nonlinear effects is much smaller than the nonlinear enhancement of the damping.

\end{abstract}

\maketitle
\section{Introduction}
\label{sec:intro}

The generation and detection of spin currents have attracted much attention in the field of spintronics. \cite{Zutic, Wolf} An effective method for detecting magnonic spin currents  is the combination of spin pumping and the inverse spin Hall effect (ISHE). Spin pumping refers to the generation of spin-polarized electron currents in a normal metal from the magnetization precession in an attached magnetic material. \cite{Tserkovnyak, Costache} These spin-polarized electron currents are transformed into conventional charge currents by the ISHE, which allows for a convenient electric detection of spin-wave spin currents.\cite{Hirsch, Saitoh-2006, Kajiwara}

After the discovery of the spin-pumping effect in ferrimagnetic insulator (yttrium iron garnet, Y$_\mathrm{3}$Fe$_\mathrm{5}$O$_\mathrm{12}$, YIG)/non-magnetic metal (platinum, Pt) heterosystems by Kajiwara et al. \cite{Kajiwara}, there was rapidly emerging interest in the investigation of these structures. \cite{Jungfleisch, Kajiwara, Chumak, Castel, Sandweg, Kurebayashi, Tashiro, Saitoh-2006} Since \YIG\ is an insulator with a bandgap of 2.85 eV \cite{Jia} no direct injection of a spin-polarized electron current into the Pt layer is possible. Thus, spin pumping in YIG/Pt structures can only be realized by exchange interaction between conduction electrons in the Pt layer and localized electrons in the YIG film. 

Spin pumping into the Pt layer transfers spin angular momentum from the YIG film thus reducing the magnetization in the YIG. This angular momentum transfer results in turn in an enhancement of the Gilbert damping of the magnetization precession. 
The magnitude of the transfer of angular momentum is independent of the ferromagnetic film thickness since spin pumping is an interface effect. However, with decreasing film thickness, the ratio between surface to volume increases and, thus, the interface character of the spin-pumping effect comes into play: the deprivation of spin angular momentum becomes notable with respect to the precession of the entire magnetization in the ferromagnetic layer. Thus, the average damping for the whole film increases with decreasing film thicknesses. It is predicted theoretically \cite{Tserkovnyak} and shown experimentally in ferromagnetic metal/normal metal heterostructures (Ni$_\mathrm{81}$Fe$_\mathrm{19}$/Pt) that the damping enhancement due to spin pumping is inversely proportional to the thickness of the ferromagnet. \cite{Nakayama-2012, Ando_JAP}

Since the direct injection of electrons from the insulator YIG into the Pt layer is not possible and spin pumping is an interface effect, an optimal interface quality is required in order to obtain a high spin- to charge current conversion efficiency. \cite{Jungfleisch-opt, Burrowes} 
Furthermore, Tashiro et al. have experimentally demonstrated that the spin mixing conductance is independent of the YIG thickness in YIG/Pt structures. \cite{Tashiro} Recently, Castel et al. reported on the YIG thickness and frequency dependence of the spin-pumping process. \cite{Castel-2013} In contrast to our investigations, they concentrate on rather thick (\textgreater 200 nm) YIG films, which are much thicker than the exchange correlation length in YIG \cite{Guslienko,Demokritov, ex-length} and thicker than the Pt thickness. Thus, the YIG film thickness dependence in the nanometer range is still not addressed till now.

In this paper, we report systematic measurements of the spin- to charge-current conversion in YIG/Pt structures as a function of the YIG film thickness from 20 nm to 300 nm. 
The Pt thickness is kept constant at 8.5 nm for all samples. 
We determine the effective damping as well as the ISHE-voltage as a function of YIG thickness and find that the thickness plays a key role. From these characteristics the spin mixing conductance and the spin Hall angle are estimated. The second part of this paper addresses microwave power dependent measurements of the ISHE-induced voltage $U_\mathrm{ISHE}$ and the ferromagnetic resonance linewidth for varying YIG film thicknesses. The occurrence of nonlinear magnon-magnon scattering processes on the widening of the linewidth as well as their influence on the spin-pumping efficiency are discussed.

\begin{figure}[t]
\includegraphics[width=0.9\columnwidth]{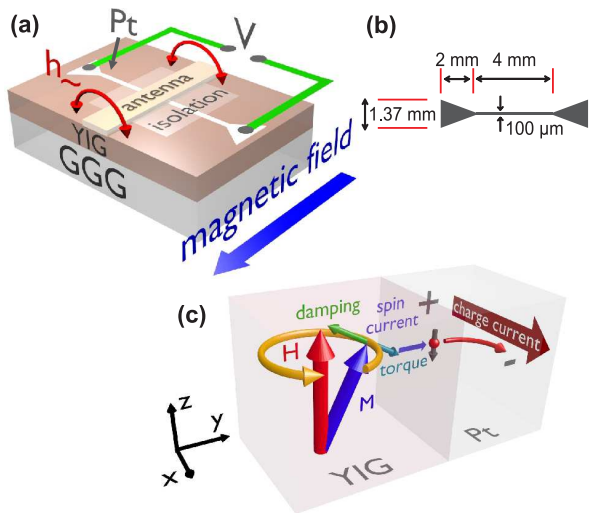}
\caption{\label{fig1} (Color online) (a) Schematic illustration of the experimental setup. (b) Dimensions of the structured Pt layer on the YIG films. The Pt layer was patterned by means of optical lithography and ion etching. (c) Scheme of combined spin-pumping process and inverse spin Hall effect.} 
\end{figure}

\begin{table}[b]
\label{table} 
\caption{ \label{table} Variation of saturation magnetization $M_\mathrm{S}$ and Gilbert damping parameter $\alpha_{0}$ as a function of the YIG film thickness. Results are obtained using a VNA-FMR measurement technique.} 
\centering  

\begin{tabular}{c  c  c} 

\hline     \hline               
$d_\mathrm{YIG}$ (nm) &  $M_\mathrm{S}$ (kA/m) &  $\alpha_\mathrm{0}$ ($\times 10^{-3}$)\\ [0.5ex] 
\hline                     
20 & 161.7 $\pm$ 0.2 & 2.169 $\pm$ 0.069  \\ 
70  & 176.4 $\pm$ 0.1 & 0.489 $\pm$ 0.007 \\
130 & 175.1 $\pm$ 0.2 & 0.430  $\pm$ 0.015 \\     
200 & 176.4 $\pm$ 0.1 & 0.162  $\pm$ 0.008 \\ 
300 & 176.5 $\pm$ 0.1 & 0.093  $\pm$ 0.007 \\ [1ex] 
\hline \hline
\end{tabular}

\end{table}

\begin{figure}[t]
\includegraphics[width=0.7\columnwidth]{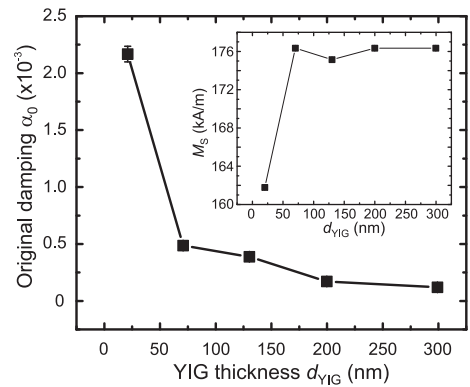}
\caption{\label{fig2} Original Gilbert damping parameter $\alpha_\mathrm{0}$ measured by VNA-FMR technique. The increased damping at low sample thicknesses is explained by an enhanced ratio between surface to volume, which results in an increased number of scattering centers and, thus, in an increased damping. The inset shows the saturation magnetization as a function of the YIG film thickness $d_\mathrm{YIG}$. The error bars are not visible in this scale.} 
\end{figure}

\section{Sample fabrication and experimental details}
\label{sec:sample}
In Fig.~\ref{fig1}(a) a schematic illustration of the investigated samples is shown. Mono-crystalline \YIG\  samples of 20, 70, 130, 200 and 300 nm thickness were deposited by means of pulsed laser deposition (PLD) from a stoichiometric target using a KrF excimer laser with 
a fluence of 2.6 J/cm$^2$ and a repetition rate of 10 Hz. \cite{Ross} In order to ensure epitaxial growth of the films, single crystalline substrates of gadolinium gallium garnet (Gd$_{3}$Ga$_{5}$O$_{12}$, GGG) in the (100) orientation were used. 
We achieved optimal deposition conditions for a substrate temperature of 650$^\circ$C $\pm$ $30^\circ$C and an oxygen pressure of 6.67$\times10^{-3}$ mbar. Afterwards, each film was annealed ex-situ at 820$^\circ$C $\pm$ $30^\circ$C by rapid thermal annealing for 300 s under a steady flow of oxygen. This improves the crystallographic order and reduces oxygen vacancies. We determined the YIG thickness by
profilometer measurements and the crystalline quality was controlled by x-ray diffraction (XRD). In order to deposit Pt onto the samples, they were transferred at atmosphere leading to possible surface adsorbates. Therefore, the YIG film surfaces were cleaned  in-situ  by a low power ion etching before the Pt deposition. \cite{Jungfleisch-opt} We used DC sputtering under an argon pressure of 1$\times$10$^{-2}$ mbar at room temperature to deposit the Pt layers. XRR measurements yielded a Pt thickness of 8.5 nm, which is identical for every sample due to the simultaneously performed Pt deposition. The Pt layer was patterned by means of optical lithography and ion etching. In order to isolate the Pt stripes from the antenna we deposited a 300 nm thick square of SU-8 photoresist on the top. A sketch of the samples and the experimental setup is shown in Fig.~\ref{fig1}(a), the dimensions of the structured Pt stripe are depicted in Fig.~\ref{fig1}(b).

In order to corroborate the quality of the fabricated YIG samples, we performed ferromagnetic resonance (FMR) measurements using a vector network analyzer (VNA).\cite{FMR}  
Since the area deposited by Pt is small compared to the entire sample size, we measure the damping  $\alpha_\mathrm{0}$ of the bare YIG by VNA (this approach results  in a small overestimate of  $\alpha_\mathrm{0}$), whereas in the spin-pumping measurement we detect the enhanced damping 
$\alpha_\mathrm{eff}$ of the Pt covered YIG films.  
The VNA-FMR results are summarized in Tab.~\ref{table} and in Fig.~\ref{fig2}. Apparently, the 20 nm sample features the largest damping of $\alpha_\mathrm{0}^\mathrm{20\,nm}=(2.169 \pm 0.069)\times 10^{-3}$. With increasing film thickness $\alpha_\mathrm{0}$ decreases to $\alpha_\mathrm{0}^\mathrm{300\,nm}=(0.093  \pm 0.007)\times 10^{-3}$. There might be two reasons for the observed behavior: (1) The quality of the thinner YIG films might be worse due to the fabrication process by PLD. (2) For smaller YIG film thicknesses, the ratio between surface to volume increases. Thus, the two-magnon scattering process at the interface is more pronounced for smaller film thicknesses and gives rise to additional damping. \cite{Sparks} 
The VNA-FMR technique yields the saturation magnetization $M_\mathrm{S}$ for the YIG samples (see inset in Fig.~\ref{fig2} and Tab.~\ref{table}). The observed values for $M_\mathrm{S}$ are larger than the bulk value, \cite{Aulock,Stancil} but in agreement with the values reported for thin films. \cite{PLD_YIG} The general trend of the film thickness dependence of $M_\mathrm{S}$ is in agreement with the one reported in  Ref.~\cite{PLD_YIG,Popova} and might be associated with a lower crystal quality after the annealing. 

The spin-pumping measurements for different YIG film thicknesses were performed in the following way. The samples were magnetized in the film plane by an external magnetic field \textbf{H}, and the magnetization dynamics was excited at a constant frequency of $f=6.8$ GHz by an Agilent E8257D microwave source. The microwave signals with powers $P_\mathrm{applied}$ of 1, 10, 20, 50, 100, 250 and 500 mW were applied to a 600 $\mu$m wide 50 Ohm-matched Cu microstrip antenna.  
While the external magnetic field was swept, the ISHE-voltage $U_\mathrm{ISHE}$ was recorded at the edges of the Pt stripe using a lock-in technique with an amplitude modulation at a frequency of 500 Hz, as well as the absorbed microwave power $P_\mathrm{abs}$. All measurements were performed at room temperature.

\section{Theoretical background}
\label{sec:theory}
The equations describing the ferromagnetic resonance, the spin pumping and the inverse spin Hall effect are provided in the following and used in the experimental part of this paper.

\subsection{Ferromagnetic resonance}
\label{FMR}
In equilibrium the magnetization \textbf{M} in a ferromagnetic material is aligned along the bias magnetic field \textbf{H}. Applying an alternating microwave magnetic field $\bf{h_\mathrm{\sim}}$ perpendicularly to the external field  \textbf{H} results in a torque on \textbf{M} and causes the magnetic moments in the sample to precess (see also Fig.~\ref{fig1}(a)). In ferromagnetic resonance (FMR) the magnetic field \textbf{H} and the precession frequency $f$ fulfill the Kittel equation \cite{Kittel} 


\begin{equation}
\label{Kittel}
f= \frac{\mu_\mathrm{0} \gamma}{2\pi} \sqrt{H_\mathrm{FMR}(H_\mathrm{FMR}+M_\mathrm{S})},
\end{equation}
where $\mu_\mathrm{0}$ is the vacuum permeability, $\gamma$ 
is the gyromagnetic ratio, $H_\mathrm{FMR}$ is the ferromagnetic resonance field and $M_\mathrm{S}$ is the saturation magnetization (experimentally obtained values of $M_\mathrm{S}$ for our samples can be found in Tab.~\ref{table}). 

The FMR linewidth $\Delta H$ (full width at half maximum) is related to the Gilbert damping parameter $\alpha$ as\cite{Nakayama-2012, Burrowes,Stancil}


\begin{equation}
\label{linewidth-alpha}
 \mu_\mathrm{0} \Delta H = 4\pi f \alpha/\gamma.
\end{equation}

\subsection{Spin pumping}
\label{SP-theory}
By attaching a thin Pt layer to a ferromagnet, the resonance linewidth is enhanced, \cite{Tserkovnyak} which accounts for an injection of a spin current from the ferromagnet into the normal metal due to the spin-pumping effect (see illustration in Fig.~\ref{fig1}(c)). In this process the magnetization precession loses spin angular momentum, which gives rise to additional damping and, thus, to an enhanced linewidth. 
The effective Gilbert damping parameter $\alpha_\mathrm{eff}$ for the YIG/Pt film is described as \cite{Nakayama-2012}

\begin{equation}
\label{alpha_eff}
\alpha_\mathrm{eff}=\alpha_\mathrm{0}+\Delta\alpha=\alpha_\mathrm{0}+\frac{g\mu_\mathrm{B}}{4\pi M_\mathrm{S} d_\mathrm{YIG}} g_\mathrm{eff}^{\uparrow \downarrow},
\end{equation}
where $\alpha_\mathrm{0}$ is the intrinsic damping of the bare YIG film, $g$ is the g-factor, $\mu_\mathrm{B}$ is the Bohr magneton, $d_\mathrm{YIG}$ is the YIG film thickness and \g\ is the real part of the effective spin mixing conductance. The effective Gilbert damping parameter $\alpha_\mathrm{eff}$ is inversely proportional to the YIG film thickness $d_\mathrm{YIG}$: with decreasing YIG thickness the linewidth and, thus, the effective damping parameter increases.

When the system is resonantly driven in the FMR condition, a spin-polarized electron current is injected from the magnetic material (YIG) into the normal metal (Pt).
In a phenomenological spin-pumping model, the DC component of the spin-current density $j_\mathrm{s}$ at the interface, injected in y-direction into the Pt layer (Fig.~\ref{fig1}(c)), can be described as \cite{Halperin, Ando_JAP, Nakayama-2012}

\begin{equation}
\label{j_s}
j_\mathrm{s}=f\int_{0}^{1/f}\frac{\hbar}{4\pi}g_\mathrm{eff}^{\uparrow \downarrow}\frac{1}{M_\mathrm{S}^2}\Big\{\mathbf{M}(t)\times\frac{d\mathbf{M}(t)}{dt} \Big\}_\mathrm{z}dt,
\end{equation}
where 
$\mathbf{M}(t)$ is the magnetization. $\{\mathbf{M}(t)\times\frac{d\mathbf{M}(t)}{dt}\}_\mathrm{z}$ is the z-component of $\{\mathbf{M}(t)\times\frac{d\mathbf{M}(t)}{dt}\}$, which is directed along  the equilibrium axis of the magnetization (see Fig.~\ref{fig1}(c)).

\begin{figure}[t]
\includegraphics[width=0.7\columnwidth]{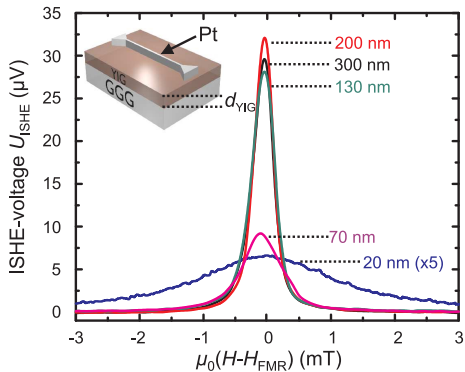}
\caption{\label{fig3} (Color online) ISHE-induced voltage $U_\mathrm{ISHE}$ as a function of the magnetic field $H$ for different YIG film thicknesses $d_\mathrm{YIG}$. Applied microwave power $P_\mathrm{applied}= 10$ mW, ISHE-voltage for the 20 nm thick sample is multiplied by a factor of 5.} 
\end{figure}

Due to spin relaxation in the normal metal (Pt) the injected spin current $j_\mathrm{s}$ decays along the Pt thickness (y-direction in Fig.~\ref{fig1}(c)) as \cite{Nakayama-2012,Ando_JAP}
\begin{equation}
\label{j_s2}
j_\mathrm{s}(y)=\frac{\mathrm{sinh}\frac{d_\mathrm{Pt}-y}{\lambda}}{\mathrm{sinh}{\frac{d_\mathrm{Pt}}{\lambda}}}j_\mathrm{s}^\mathrm{0},
\end{equation}
where $\lambda$ is the spin-diffusion length in the Pt layer.
From Eq.~(\ref{j_s}) one can deduce the spin-current density at the interface ($y=0$) \cite{Ando_JAP}
\begin{equation}
\label{j_s3}
j_\mathrm{s}^\mathrm{0}=\frac{g_\mathrm{eff}^{\uparrow \downarrow}\gamma^2 (\mu_\mathrm{0} h_\mathrm{\sim})^2 \hbar (\mu_\mathrm{0} M_\mathrm{S}\gamma + \sqrt{(\mu_\mathrm{0} M_\mathrm{S} \gamma)^2 + 16(\pi f)^2})}{8\pi \alpha_\mathrm{eff}^2((\mu_\mathrm{0} M_\mathrm{S}\gamma)^2+16(\pi f)^2)}.
\end{equation}
Since $j_\mathrm{s}^\mathrm{0}$ is inversely proportional to $\alpha_\mathrm{eff}^2$ and $\alpha_\mathrm{eff}$ depends inversely on $d_\mathrm{YIG}$ (Eq.~(\ref{alpha_eff})), the spin-current density at the interface $j_\mathrm{s}^\mathrm{0}$ increases with increasing YIG film thickness $d_\mathrm{YIG}$.

\subsection{Inverse spin Hall effect}
\label{ISHE-theory}

The Pt layer acts as a spin-current detector and transforms the spin-polarized electron current injected due to the spin-pumping effect into an electrical charge current by means of the ISHE (see Fig.~\ref{fig1}(c)) as\cite{Saitoh-2006, Kajiwara, Nakayama-2012, Ando_JAP, Jungfleisch}

\begin{equation}
\label{j_c}
\mathbf{j_\mathrm{c}}=\theta_\mathrm{ISHE}\frac{2e}{\hbar}\mathbf{j_s}\times \mathbf{\sigma},
\end{equation}
where $\theta_\mathrm{ISHE}$, $e$,  $\sigma$ denote the spin Hall angle, the electron's elementary charge and the spin-polarization vector, respectively. 
Averaging the charge-current density over the Pt thickness and taking into account Eqs.~(\ref{j_s}) -- (\ref{j_c})  yields
\begin{equation}
\label{j_c_av}
\bar{j_\mathrm{c}}=\frac{1}{d_\mathrm{Pt}}\int_\mathrm{0}^\mathrm{d_\mathrm{Pt}}j_\mathrm{c}(y)dy=\theta_\mathrm{ISHE} \frac{\lambda}{d_\mathrm{Pt}}\frac{2e}{\hbar}\mathrm{tanh}\big(\frac{d_\mathrm{Pt}}{2\lambda}\big)j_\mathrm{s}^\mathrm{0}.
\end{equation}

Taking into account Eqs.~(\ref{alpha_eff}), (\ref{j_s3}) and (\ref{j_c_av}) we calculate the theoretically expected behavior of $I_\mathrm{ISHE}=A \bar{j_\mathrm{c}}$, where $A$ is the cross section of the Pt layer.
Ohm's law connects the ISHE-voltage $U_\mathrm{ISHE}$ with the ISHE-current $I_\mathrm{ISHE}$ via $U_\mathrm{ISHE}$=$I_\mathrm{ISHE}\cdot R$, where $R$ is the electric resistance of the Pt layer. $R$ varies between 1450 $\Omega$ and 1850 $\Omega$ for the different samples.

\begin{figure}[b]
\includegraphics[width=0.7\columnwidth]{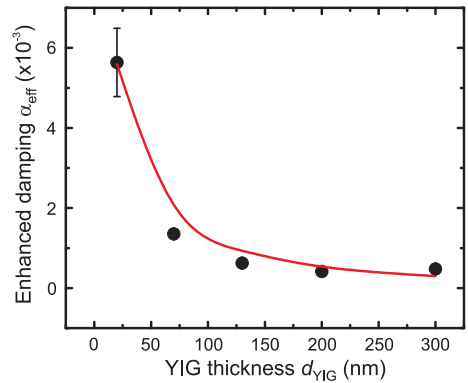}
\caption{\label{fig4} (Color online) Enhanced damping parameter $\alpha_\mathrm{eff}$ of the YIG/Pt samples obtained by spin-pumping measurements. The red solid curve shows a fit to Eq.~(\ref{alpha_eff}) taking the FMR measured values for $M_\mathrm{S}$ and a constant value for \g\ into account. $P_\mathrm{applied}$= 1 mW. The error bars for the measurement points at higher sample thicknesses are not visible in this scale.} 
\end{figure}

\section{YIG film thickness dependence of the spin-pumping effect detected by the ISHE}
\label{sec:thick}

In Fig.~\ref{fig3} the magnetic field dependence of the generated ISHE-voltage $U_\mathrm{ISHE}$ as a function of the YIG film thickness is shown. Clearly, the maximal voltage $U_\mathrm{ISHE}$ at the resonance field $H_\mathrm{FMR}$ and the FMR linewidth $\Delta H$ vary with the YIG film thickness. The general trend shows, that the thinner the sample the smaller is the magnitude of the observed voltage $U_\mathrm{ISHE}$. At the same time the FMR linewidth increases with decreasing YIG film thickness.

In the following the ISHE-voltage generated by spin pumping is investigated as a function of the YIG film thickness. For these investigations we have chosen a rather small exciting microwave power of 1 mW. Thus, nonlinear effects like the FMR linewidth broadening due to nonlinear multi-magnon processes can be excluded (such processes will be discussed in Sec.~\ref{sec:non-linear}).
Sec.~\ref{subsec:alpha} covers the YIG thickness dependent variation of the enhanced damping parameter $\alpha_\mathrm{eff}$. From these measurements the spin mixing conductance \g\ is deduced.
In Sec.~\ref{subsec:thick} we focus on the maximal ISHE-voltage driven by spin pumping as a function of the YIG film thickness. Finally, the spin Hall angle $\theta_\mathrm{ISHE}$ is determined. 

\begin{figure}[t]
\includegraphics[width=0.7\columnwidth]{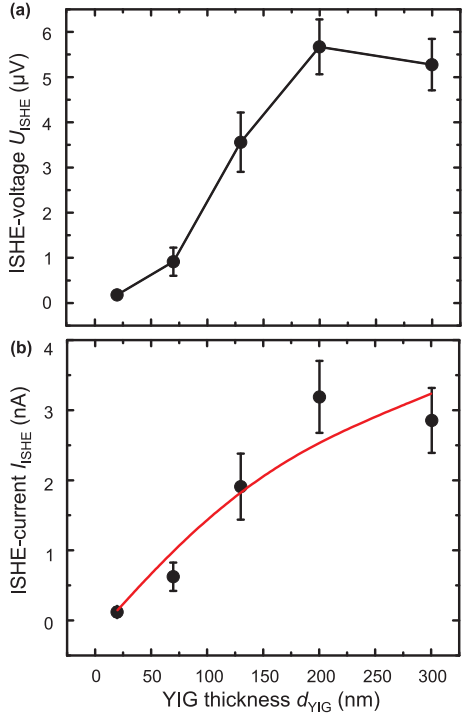}
\caption{\label{fig5} (Color online) (a) ISHE-voltage $U_\mathrm{ISHE}$ as a function of the YIG film thickness $d_\mathrm{YIG}$. The black line is a linear interpolation as a guide to the eye. (b) Corresponding thickness dependent charge current $I_\mathrm{ISHE}$. The red curve shows a fit to Eqs.~(\ref{j_s3}), (\ref{j_c}), (\ref{j_c_av}) with the parameters \g~=~(7.43$ \pm$0.36)$ \times 10^{18}$ m$^{-2}$ and $\theta_\mathrm{ISHE}= 0.009 \pm 0.0008$. The applied microwave power used is $P_\mathrm{applied}=1$ mW.} 
\end{figure}

\subsection{YIG film parameters as a function of the YIG film thickness}
\label{subsec:alpha}

As described in Sec.~\ref{SP-theory}, the damping parameter is enhanced when a Pt layer is deposited onto the YIG film. This enhancement is investigated as a function of the YIG film thickness: the effective Gilbert damping parameter $\alpha_\mathrm{eff}$ (see Eq.~(\ref{alpha_eff})) is obtained from a Lorentzian fit to the experimental data depicted in Fig.~\ref{fig3} and Eq.~(\ref{linewidth-alpha}). The result is shown in Fig.~\ref{fig4}. With decreasing YIG film thickness the linewidth and, thus, the effective damping $\alpha_\mathrm{eff}$ increases. This behavior is theoretically expected: according to Eq.~(\ref{alpha_eff}) $\alpha_\mathrm{eff}$ is inversely proportional to $d_\mathrm{YIG}$. Since the Pt film is grown onto all YIG samples simultaneously, the spin mixing conductance \g\ at the interface is considered to be constant for all samples. \cite{Tashiro} Assuming \g\ as constant and taking the saturation magnetization $M_\mathrm{S}$ obtained by VNA-FMR measurements (see Fig.~\ref{fig2} and Tab.~\ref{table}) into account, a fit to Eq.~(\ref{alpha_eff}) yields \g~=~(7.43$ \pm$0.36)$ \times 10^{18}$ m$^{-2}$. The fit is depicted as a red solid line in Fig.~\ref{fig4}.

\subsection{YIG thickness dependence of the ISHE-voltage driven by spin pumping}
\label{subsec:thick}

Fig.~\ref{fig5}(a) shows the maximum voltage $U_\mathrm{ISHE}$ at the resonance field $H_\mathrm{FMR}$ as a function of the YIG film thickness. $U_\mathrm{ISHE}$ increases up to a YIG film thickness of around 200 nm when it starts to saturate (in the case of an applied microwave power of P$_\mathrm{applied}=1$ mW).
The corresponding charge current $I_\mathrm{ISHE}$ is shown in Fig.~\ref{fig5}(b). The observed thickness dependent behavior is in agreement with the one reported for Ni$_\mathrm{81}$Fe$_\mathrm{19}$/Pt \cite{Nakayama-2012} and for \YIG/Pt. \cite{Tashiro} With increasing YIG film thickness the generated ISHE-current increases and tends to saturate at thicknesses near 200 nm (Fig.~\ref{fig5}(b)). According to Eq.~(\ref{alpha_eff}), (\ref{j_s3}) and (\ref{j_c_av}) it is $I_\mathrm{ISHE} \propto  j_\mathrm{s}^0 \propto 1/\alpha_\mathrm{eff}^2 \propto (\alpha_\mathrm{0} + c/d_\mathrm{YIG})^{-2}$, where $c$ is a constant. Therefore, the ISHE-current $I_\mathrm{ISHE}$ increases with increasing YIG film thickness $d_\mathrm{YIG}$ and goes into saturation at a certain YIG thickness.

\begin{figure}[b]
\includegraphics[width=0.8\columnwidth]{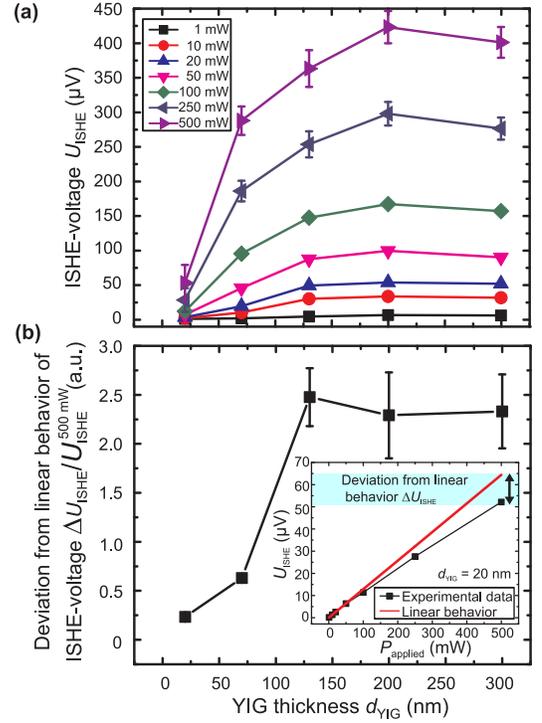}
\caption{\label{fig7} (Color online) (a) YIG thickness dependence of the ISHE-voltage driven by spin pumping for microwave powers in the range between 1 and 500 mW. The general thickness dependent behavior is independent of the applied microwave power. The error bars for the measurement at lower microwave powers are not visible in this scale. (b) Deviation of the ISHE-voltage from the linear behavior with respect to the measured voltage $U_\mathrm{ISHE}^\mathrm{500 mW}$. The inset shows experimental data for a YIG film thickness $d_\mathrm{YIG}=20$ nm and the theoretically expected curve. The error bars of the 20 nm and the 70 nm samples are not visible in this scale.}
\end{figure}

From Eqs.~(\ref{alpha_eff}), (\ref{j_s3}) and (\ref{j_c_av}) we determine the expected behavior of $I_\mathrm{ISHE}=A \bar{j_\mathrm{c}}$ and compare it with our experimental data. In order to do so, the measured values for $M_\mathrm{S}$ (see Tab.~\ref{table}), the original damping parameter $\alpha_\mathrm{0}$ determined by VNA-FMR measurements at 1 mW (see Tab.~\ref{table}) and the enhanced damping parameter $\alpha_\mathrm{eff}$ obtained by spin-pumping measurements at a microwave power of 1 mW (see Fig.~\ref{fig4}) are used. The Pt layer thickness is $d_\mathrm{Pt}= 8.5$ nm and the microwave magnetic field is determined to be $h_\sim = 3.2$~A/m for an applied microwave power of 1 mW using an analytical expression. \cite{Chumakov} The spin-diffusion length in Pt is taken from literature as $\lambda = 10$ nm \cite{Mosendz, Kurt} and the damping parameter is assumed to be constant as $\alpha_{0}= 6.68\times 10^{-4}$, which is the average of the measured values of $\alpha_\mathrm{0}$. The fit is shown as a red solid line in Fig.~\ref{fig5}(b). We find a spin Hall angle of $\theta_\mathrm{ISHE}= 0.009 \pm 0.0008$, which is in agreement with literature values varying in a range of 0.0037 - 0.08. \cite{Mosendz,Ando_PRL,Kimura} Using the fit we estimate the saturation value of the generated current. Although we observe a transition to saturation at sample thicknesses of 200 -- 300 nm, we find that according to our fit, 90\% of the estimated saturation level of 5 nA is reached at a sample thickness of 1.2 $\mu$m.

\begin{figure}[t!]
\includegraphics[width=0.85\columnwidth]{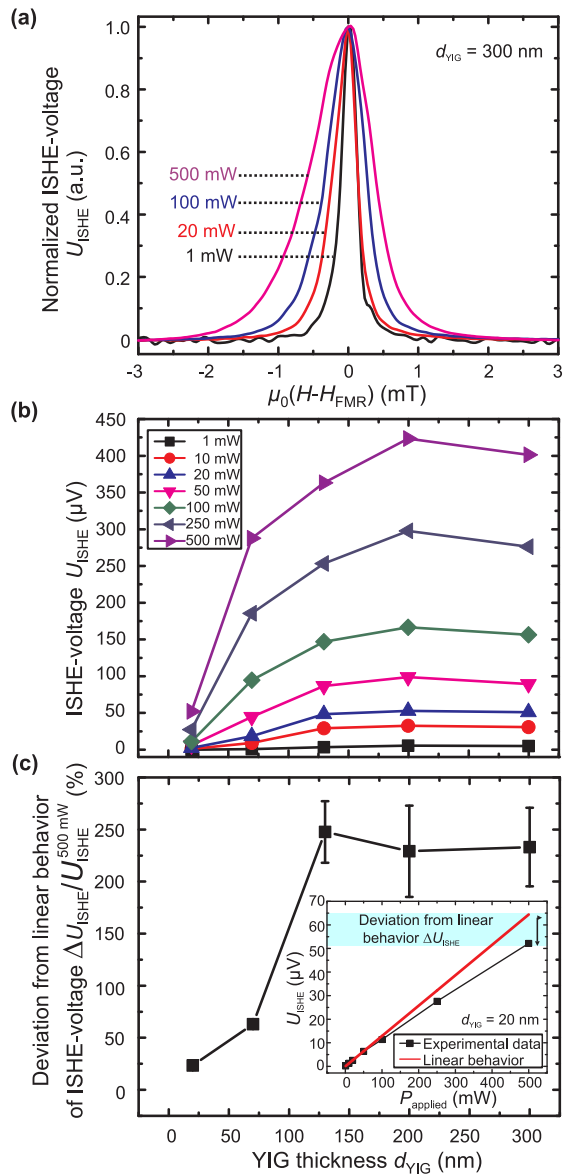}
\caption{\label{fig8} (Color online) (a) Illustration of the linewidth broadening at higher excitation powers. The normalized ISHE-voltage spectra are shown as a function of the magnetic field $H$ for different excitation powers. Sample thickness: 300 nm. (b) Power dependent measurement of the damping parameter $\alpha_\mathrm{eff}$ for different YIG film thicknesses $d_\mathrm{YIG}$ obtained by a Lorentzian fit to the ISHE-voltage signal. The error bars are omitted in order to provide a better readability of the viewgraph. (c) Nonlinear damping enhancement $(\alpha_\mathrm{eff}^\mathrm{500 mW}-\alpha_\mathrm{eff}^\mathrm{1 mW})/\alpha_\mathrm{eff}^\mathrm{1 mW}$ as a function of the YIG film thickness $d_\mathrm{YIG}$. Due to a reduced number of scattering channels to other spin-wave modes for film thicknesses below 70 nm, the damping is only enhanced for thicker YIG films with increasing applied microwave powers. The error bars of the 200 nm and the 300 nm samples are not visible on this scale.}
\end{figure}

\begin{figure*}[t]
\includegraphics[width=1.75\columnwidth]{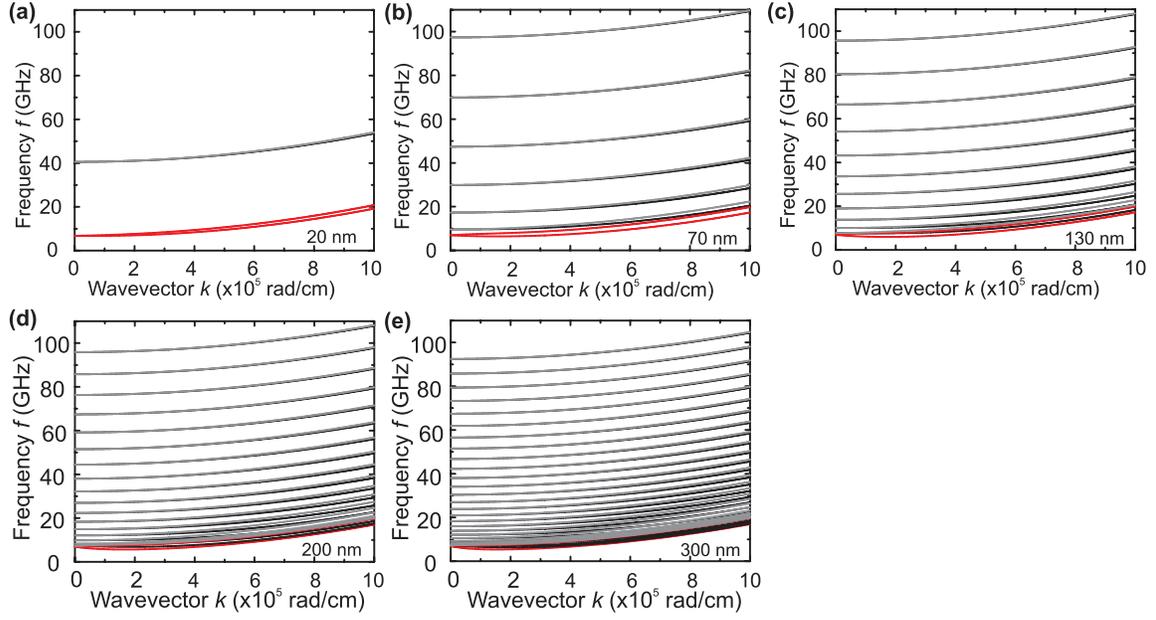}
\caption{\label{fig9} (Color online) Dispersion relations calculated for each sample thickness taking into account the measured values of the saturation magnetization $M_\mathrm{S}$ (see Tab.~\ref{table}). Backward volume magnetostatic spin-wave modes as well as magnetostatic surface spin-wave modes  (in red) and the first perpendicular standing thickness spin-wave modes are depicted (in black and gray). (a)--(e) show the dispersion relations for the investigated sample thicknesses of 20 nm -- 300 nm.}
\end{figure*}

\section{Influence of nonlinear effects on the spin-pumping process for varying YIG film thicknesses}
\label{sec:non-linear}

In order to investigate nonlinear effects on the spin-pumping effect for varying YIG film thicknesses, we performed microwave power dependent measurements of the ISHE-voltage $U_\mathrm{ISHE}$ as function of the film thickness $d_\mathrm{YIG}$.
For higher microwave powers in the range of 1 mW to 500 mW we observe the same thickness-dependent behavior of the ISHE-voltage as in the linear case ($P_\mathrm{applied} = 1$ mW, discussed in Sec.~\ref{subsec:thick}): Near 200 nm $U_\mathrm{ISHE}$ starts to saturate independently of the applied microwave power, as it is shown in Fig.~\ref{fig7}(a). 
Furthermore, it is clearly visible from Fig.~\ref{fig7}(a) that for a constant film thickness the spin pumping driven ISHE-voltage increases with increasing applied microwave power. At high microwave powers the voltage does not grow linearly and saturates. Fig.~\ref{fig7}(b) shows the deviation of the ISHE-voltage $\Delta U_\mathrm{ISHE}$ from the linear behavior with respect to the measured value of $U_\mathrm{ISHE}^\mathrm{500 mW}$ at the excitation power $P_\mathrm{applied} = $ 500~mW. In order to obtain the relation between $U_\mathrm{ISHE}$ and $P_\mathrm{applied}$ for each YIG film thickness $d_\mathrm{YIG}$ the low power regime up to 20 mW is fitted by a linear curve and extrapolated to 500 mW. The inset in Fig.~\ref{fig7}(b) shows the corresponding viewgraph for the case of the 20 nm thick sample. As it is visible from Fig.~\ref{fig7}(b), the deviation from the linear behavior is drastically enhanced for larger YIG thicknesses. For the thin 20 nm and 70 nm samples we observe an almost linear behavior between $U_\mathrm{ISHE}$ and $P_\mathrm{applied}$ over the entire microwave power range, whereas for the thicker samples the estimated linear behavior and the observed nonlinear behavior differ approximately by a factor of 2.5 (Fig.~\ref{fig7}(b)). 
We observe an increase of the ISHE-voltage as well as an broadening of the FMR linewidth with increasing microwave power.
In Fig.~\ref{fig8}(a) the normalized ISHE-voltage $U_\mathrm{ISHE}$ as function of the external magnetic field $H$ is shown for different microwave powers $P_\mathrm{applied}$ in the range of 1 mW to 500 mW (YIG film thickness $d_\mathrm{YIG}= 300$ nm). The linewidth tends to be asymmetric at higher microwave powers. The shoulder at lower magnetic field is widened in comparison to the shoulder at higher fields. The reason for this asymmetry might be due to the formation of a foldover effect, \cite{Gui,Ando_bistable} due to nonlinear damping or a nonlinear frequency shift. \cite{Demidov_non-linear, Mills}

The results of the damping parameter $\alpha_\mathrm{eff}$ obtained by microwave power dependent spin-pumping measurements are depicted in Fig.~\ref{fig8}(b). It can be seen, that with increasing excitation power the Gilbert damping for thicker YIG films is drastically increased. To present  this result more clearly the nonlinear damping enhancement  $(\alpha_\mathrm{eff}^\mathrm{500 mW}-\alpha_\mathrm{eff}^\mathrm{1 mW})/\alpha_\mathrm{eff}^\mathrm{1 mW}$ is shown in Fig.~\ref{fig8}(c). The damping parameter at a sample thickness of 20 nm $\alpha_\mathrm{eff}^\mathrm{20 nm}$ is almost unaffected by a nonlinear broadening at high microwave powers. With increasing film thickness the original damping $\alpha_\mathrm{eff}^\mathrm{1 mW}$  at  $P_\mathrm{applied}= 1$ mW increases by a factor of around 3 at $P_\mathrm{applied}=500$ mW. This factor is very close to the value of the deviation of the ISHE-voltage from the linear behavior (Fig.~\ref{fig7}(b)).

This behavior can be attributed to the enhanced probability of nonlinear multi-magnon processes at larger sample thicknesses: 
In order to understand this, a fundamental understanding of the restrictions for multi-magnon scattering processes can be derived from the energy and momentum conservation laws:
\begin{equation}
\label{energy}
\sum_\mathrm{i}^\mathrm{N} \hbar \omega_\mathrm{i} = \sum_\mathrm{j}^\mathrm{M} \hbar \omega_\mathrm{j} \mathrm{\, \, and \, \, } \sum_\mathrm{i}^\mathrm{N} \hbar \mathbf{k}_\mathrm{i} = \sum_\mathrm{j}^\mathrm{M} \hbar\mathbf{k}_\mathrm{j},
\end{equation}
where the left/right sum of the equations runs over the initial/final magnons with indices i/j which exist before/after the scattering process, respectively. \cite{Schultheiss,Sebastian,BEC_nature} The most probable scattering mechanism in our case is the four-magnon scattering process with $N=2$ and $M=2$. \cite{BEC_nature} In Eq.~(\ref{energy}) the wavevector $\mathbf{k}_\mathrm{i/j}$ and the frequency $\omega_\mathrm{i/j}$ are connected by the dispersion relation $2\pi f_\mathrm{i/j}(\mathbf{k}_\mathrm{i/j})=\omega_\mathrm{i/j}(\mathbf{k}_\mathrm{i/j})$. The calculated dispersion relations are shown in Fig.~\ref{fig9} (backward volume magnetostatic spin-wave modes with a propagation angle $\angle(\mathbf{H},\mathbf{k})= 0^\circ$ as well as magnetostatic surface spin-wave modes $\angle(\mathbf{H},\mathbf{k})= 90^\circ$). \cite{DamonEshbach} For this purpose, the measured values of $M_\mathrm{S}$ (Tab.~\ref{table}) for each sample are used. 
In the case of the 20 nm sample thickness, the first perpendicular standing spin-wave mode (thickness mode) lies above 40 GHz, the second above 120 GHz. Thus, the nonlinear scattering probability obeying the energy- and momentum conservation is largely reduced. 
This means magnons cannot find a proper scattering partner and, thus, multi-magnon processes are prohibited or at least largely suppressed.
With increasing film thickness the number of standing spin-wave modes increases and, thus, the scattering probability grows. As a result, the scattering of spin waves from the initially excited uniform precession (FMR) to other modes is allowed and the relaxation of the original FMR mode is enhanced. Thus, the damping increases and we observe a broadening of the linewidth, which is equivalent to an enhanced Gilbert damping parameter $\alpha_\mathrm{eff}$ at higher YIG film thicknesses (see Fig.~\ref{fig8}).

\begin{figure}[t]
\includegraphics[width=0.8\columnwidth]{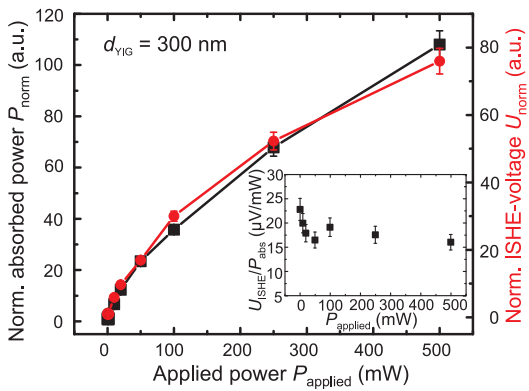}
\caption{\label{fig6} (Color online) Normalized absorbed power $P_\mathrm{norm}=P_\mathrm{abs}/P_\mathrm{abs}^{P_\mathrm{applied}= 1  \mathrm{mW}}$ (black squares) and normalized ISHE-voltage $U_\mathrm{norm}=U_\mathrm{ISHE}/U_\mathrm{ISHE}^{P_\mathrm{applied}=1 \mathrm{mW}}$ (red dots) for varying microwave powers $P_\mathrm{applied}$. The inset illustrates the independence of the spin-pumping efficiency $U_\mathrm{ISHE}/P_\mathrm{abs} $ on $P_\mathrm{applied}$. YIG thickness illustrated: 300 nm. Error bars of the low power measurements are not visible in this scale.}
\end{figure}

In order to investigate how the spin-pumping efficiency is affected by the applied microwave power, we measure simultaneously the generated ISHE-voltage $U_\mathrm{ISHE}$ and the transmitted ($P_\mathrm{trans}$), as well as the reflected ($P_\mathrm{refl}$) microwave power, which enables us to determine the absorbed microwave power $P_\mathrm{abs}=P_\mathrm{applied}-(P_\mathrm{trans}+P_\mathrm{refl})$. \cite{Jungfleisch-opt}
Since the 300 nm sample exhibits a strong non-linearity (large deviation from the linear behavior (Fig.~\ref{fig7}) and large nonlinear linewidth enhancement (Fig.~\ref{fig8})), we analyze this sample thickness. In Fig.~\ref{fig6} the normalized absorbed microwave power $P_\mathrm{norm}=P_\mathrm{abs}/P_\mathrm{abs}^{P_\mathrm{applied}=1 \mathrm{ mW}}$ and the normalized ISHE-voltage in resonance  $U_\mathrm{norm}=U_\mathrm{ISHE}/U_\mathrm{ISHE}^{P_\mathrm{applied}=1 \mathrm{ mW}}$ are shown as a function of the applied power $P_\mathrm{applied}$. Both curves tend to saturate at high microwave powers above 100 mW. The absorbed microwave power increases by a factor of 110 for applied microwave powers in the range between 1 and 500 mW, whereas the generated voltage increases by a factor of 80. The spin-pumping efficiency $U_\mathrm{ISHE}/P_\mathrm{abs}$ (see inset in Fig.~\ref{fig6}) varies within a range of 30\% for the different microwave powers $P_\mathrm{applied}$ without clear trend. Since the 300 nm thick film shows a nonlinear deviation of the ISHE-voltage by a factor of 2.3 (Fig.~\ref{fig7}(b)) and the damping is enhanced by a factor of 3 in the same range of $P_\mathrm{applied}$ (Fig.~\ref{fig8}(c)), we conclude that the spin-pumping process is only weakly dependent on the magnitude of the applied microwave power (see inset in Fig.~\ref{fig6}). In our previous studies reported in Ref.~\cite{Jungfleisch, Chumak} we show that secondary magnons generated in a process of multi-magnon scattering contribute to the spin-pumping process and, thus, the spin-pumping efficiency does not depend on the applied microwave power.

\section{Summary}
\label{sec:sum}

The \YIG\ thickness dependence of the spin-pumping effect detected by the ISHE has been investigated quantitatively. It is shown that the effective Gilbert damping parameter of the the YIG/Pt samples is enhanced for smaller YIG film thicknesses, which is attributed to an increase of the ratio between surface to volume and, thus, to the interface character of the spin-pumping effect. We observe a theoretically expected increase of the ISHE-voltage with increasing YIG film thickness tending to saturate above thicknesses near 200 -- 300 nm. The spin mixing conductance  \g~=~(7.43$ \pm$0.36)$ \times 10^{18}$ m$^{-2}$ as well as the spin Hall angle $\theta_\mathrm{ISHE} = 0.009 \pm 0.0008$ are calculated and are found to be in agreement with values reported in the literature for our materials.

The microwave power dependent measurements reveal the occurrence of nonlinear effects for the different YIG film thicknesses: for low powers, the induced voltage grows linearly with the power. At high powers, we observe a saturation of the ISHE-voltage $U_\mathrm{ISHE}$ and a deviation by a factor of 2.5 from the linear behavior. The  microwave power dependent investigations of the Gilbert damping parameter by spin pumping show an enhancement by a factor of 3 at high sample thicknesses due to nonlinear effects. This enhancement of the damping is due to nonlinear scattering processes representing an additional damping channel which absorbs energy from the originally excited FMR. We have shown that the smaller the sample thickness, the less dense is the spin-wave spectrum and, thus, the less nonlinear scattering channels exist. Hence, the smallest investigated sample thicknesses (20 and 70 nm) exhibit a small deviation of the ISHE-voltage from the linear behavior and  a largely reduced enhancement of the damping parameter at high excitation powers. Furthermore, we have found that the variation of the spin-pumping efficiencies for thick YIG samples which show strongly nonlinear effects is much smaller than the nonlinear enhancement of the damping. This is attributed to secondary magnons generated in a process of multi-magnon scattering that contribute to the spin pumping. It is shown, that even for thick samples (300 nm) the spin-pumping efficiency is only weakly dependent on the applied microwave power and varies only within a range of 30\% for the different microwave powers without a clear trend.

Our findings provide a guideline to design and create efficient magnon- to charge current converters. Furthermore, the results are also substantial for the reversed effects: the excitation of spin waves in thin YIG/Pt bilayers by the direct spin Hall effect and the spin-transfer torque effect. \cite{Berger}

\section{Acknowledgments}
\label{sec:ack}

We thank G.E.W.~Bauer and V.I.~Vasyuchka for valuable discussions. Financial support by the Deutsche Forschungsgemeinschaft within the project CH 1037/1-1 are gratefully acknowledged. AK would like to thank the Graduate School of Excellence Materials Science in Mainz (MAINZ) GSC 266. CAR, MCO and DHK acknowledge support from the National Science Foundation. Shared experimental facilities supported by NSF MRSEC award DMR-0819762 were used.


\begin{thebibliography}{19}
\bibitem{ex-length}The exchange correlation length is given by $l_\mathrm{ex}=\sqrt{A/(2\pi M_\mathrm{S}^2)}$, where $A$ is the exchange constant and $M_\mathrm{S}$ is the saturation magnetization. With $A=3.6\times 10^{-7}$ erg/cm we obtain an exchange length of $l_\mathrm{ex}\approx 17$ nm for YIG.
\end{thebibliography}
\end{document}